# GENERALIZED FUNDAMENTAL SOLUTIONS FOR UNSTEADY VISCOUS FLOWS

## JIAN-JUN SHU AND ALLEN T. CHWANG

School of Mechanical & Aerospace Engineering
Nanyang Technological University
50 Nanyang Avenue, Singapore 639798
e-mail: mjjshu@ntu.edu.sg, http://www.ntu.edu.sg/home/mjjshu

**Abstract.** A number of new closed-form fundamental solutions for the generalized unsteady Oseen and Stokes flows associated with arbitrary time-dependent translational and rotational motions have been developed. These solutions are decomposed into two parts corresponding to a longitudinal wave and a transversal wave. As examples of application, the hydrodynamic force acting on a sphere or on a circular cylinder translating in an unsteady flow field at low Reynolds numbers are calculated using the new generalized fundamental solutions. The results for a rotating viscous flow past an impulsively moving sphere and an impulsively moving circular cylinder are entirely new.

## 1  INTRODUCTION

In investigating flows at low Reynolds numbers, it has long been customary to linearize the Navier-Stokes equations in order to obviate a prohibitively difficult problem of obtaining complete, analytical solutions. Stokes [1] investigated the case of a parallel flow past a sphere and proposed the oldest known linearization. He acknowledged that, under his linearization, it was impossible to find a solution for a two-dimensional viscous flow past a finite body, a conclusion now known as the "Stokes paradox". Oseen [2] included a translational inertia term in the Navier-Stokes equations and gave an improvement of the Stokes linearization.

A useful method for solving such linearized flows is the singularity method, in which the solution is expressed in terms of discrete or continuous distributions of fundamental singularities. The success of the method depends mainly upon the choice of the correct type of fundamental singularities and their spatial distributions. For inviscid flows, the fundamental singularities such as sources, vortices, dipoles and their usage for complicated flow situations have been thoroughly studied. For steady viscous flows, Chwang & Wu [3] introduced a set of new fundamental solutions called the Stokesons, rotons and stressons, which have been further applied to a wide variety of flow problems. For unsteady viscous flows, Pozrikidis [4] derived expressions for an oscillating Stokeslet and dipole to study the viscous oscillatory flow past a spheroid. Price & Tan [5] gave a convolution integral formulation for the transient Oseenlets associated with a body maneuvering in a viscous fluid. Further references can be found on some special cases, such as a Laplacian representation on an oscillating Stokeslet and a concise presentation on a purely translating Oseenlet with a prescribed constant velocity.

The drag on a body in transient motion has been of long-standing interest. Sano [6] obtained a long-time representation for the force on a sphere in response to an impulsively started flow at a small but finite Reynolds number using the method of matched asymptotic expansions. Nakanishi *et al*. [7] treated a two-dimensional version of the problem. Lovalenti & Brady [8,9] extended Sano's result to a step change in the free-stream velocity using a reciprocal theorem. Tanzosh & Stone [10] studied the steady motion of a rigid particle in a rotating viscous flow using an integral equation approach. A keen insight in considering the long-time decay of the drag on a body as steady state is approached for finite Reynolds numbers has been addressed. However, the generalized fundamental solutions for an arbitrary, temporal domain still remain difficult to obtain. In this paper, we shall consider time-dependent linearized viscous flows, taking both translational and rotational motions into account. In the work that follows, it is demonstrated that the derivation of the net force on a body is especially simple by employing new generalized fundamental solutions to construct exact solutions. Although this paper is devoted to studying drag forces on a sphere or a circular cylinder translating impulsively from rest in a rotating viscous flow, these are regarded as two prototypes of a general class of problems. It is understood that the essential features of the formulation and solutions may readily be applied to other general cases [11].

## 2 GOVERNING EQUATIONS

Let us consider an unsteady flow with translational velocity $\mathbf{U}^*(t)$ and angular velocity $\mathbf{\Omega}^*(t)$ past a stationary body. The flow starts from rest, *i.e.* $\mathbf{U}^*(0)=\mathbf{O}$, $\mathbf{\Omega}^*(0)=\mathbf{O}$. Let us nondimensionalize time by $\dfrac{L}{U R_e}$, distance by $\dfrac{L}{R_e}$, velocity by $U$, pressure by $\rho U^2$ and $\mathbf{U}(t)=\dfrac{\mathbf{U}^*(t)}{U}$, $\mathbf{\Omega}(t)=\dfrac{\mathbf{\Omega}^*(t)L}{U R_e}$, where $L$ and $U$ are the characteristic length and speed, $R_e=\dfrac{\rho U L}{\mu}$ is the Reynolds number, $\rho$ and $\mu$ are the density and viscosity of the fluid. The unsteady flow is governed by the dimensionless Navier-Stokes equations

$$\nabla \bullet \mathbf{V} = 0, \qquad (1)$$

$$\frac{\partial \mathbf{V}}{\partial t} + \mathbf{V}\bullet\nabla\mathbf{V} = -\nabla P + \nabla^2 \mathbf{V} + \frac{d\mathbf{U}}{dt} + \frac{d\mathbf{\Omega}}{dt}\times\mathbf{x} - \mathbf{\Omega}\times(\mathbf{\Omega}\times\mathbf{x}) + 2\mathbf{\Omega}\times\mathbf{V} + \mathbf{F}(t,\mathbf{x}), \qquad (2)$$

where $\mathbf{V}$, $P$, $\mathbf{F}(t,\mathbf{x})$ and $\mathbf{x}$ denote the non-dimensional velocity vector, pressure, external body force strength and position vector measured in a Cartesian coordinate system $(\mathbf{e}_1,\mathbf{e}_2,\mathbf{e}_3)$ with the origin located at the instantaneous center of the body. We consider the disturbed fluid velocity $\mathbf{u}$ and the disturbed pressure $p$ in the fluid as the basic unknowns. Thus, letting $\mathbf{V}=\mathbf{u}+\mathbf{U}+\mathbf{\Omega}\times\mathbf{x}$, the linearized equations (1) and (2) become

$$\nabla \bullet \mathbf{u} = 0, \qquad (3)$$

$$\frac{\partial \mathbf{u}}{\partial t} + \mathbf{U}\bullet\nabla\mathbf{u} - \nabla\times[(\mathbf{\Omega}\times\mathbf{x})\times\mathbf{u}] = -\nabla p + \nabla^2 \mathbf{u} + \mathbf{F}(t,\mathbf{x}). \qquad (4)$$

We identify the first term on the left-hand side of (4) as the unsteady inertial term, the second term as the convective inertial term and the third term as the Coriolis term. If $\mathbf{U}(t)=\mathbf{U}_0$ (a constant) and $\mathbf{\Omega}(t) = \mathbf{O}$ for all $t > 0$, equation (4) reduces to the well-known unsteady Oseen equation [2],

$$\frac{\partial \mathbf{u}}{\partial t} + \mathbf{U}_0 \bullet \nabla \mathbf{u} = -\nabla p + \nabla^2 \mathbf{u} + \mathbf{F}(t, \mathbf{x}). \tag{5}$$

If $\mathbf{U}_0$ vanishes, equation (5) further reduces to the well-known unsteady Stokes equation [1],

$$\frac{\partial \mathbf{u}}{\partial t} = -\nabla p + \nabla^2 \mathbf{u} + \mathbf{F}(t, \mathbf{x}).$$

## 3  GENERALIZED FUNDAMENTAL SOLUTIONS

For a given body and prescribed motion, equations (3) and (4) have a unique solution in a Euclidean temporal and spatial domain in terms of fundamental solutions $\mathbf{u}(t, \mathbf{x}|t_0, \mathbf{x}_0)$ and $p(t, \mathbf{x}|t_0, \mathbf{x}_0)$, $[\mathbf{u}, p](t, \mathbf{x}) = \int [\mathbf{u}, p](t, \mathbf{x}|t_0, \mathbf{x}_0) \mathbf{F}(t_0, \mathbf{x}_0) dt_0 \, d\mathbf{x}_0$. The fundamental solutions $\mathbf{u}(t, \mathbf{x}|t_0, \mathbf{x}_0)$ and $p(t, \mathbf{x}|t_0, \mathbf{x}_0)$ with respect to the singular point $\{t_0, \mathbf{x}_0\}$ satisfy equations (3) and (4) with $\mathbf{F}(t, \mathbf{x})$ replaced by $\mathbf{F}\delta(t - t_0)\delta(\mathbf{x} - \mathbf{x}_0)$, where $\mathbf{F}$ is a constant force vector and $\delta(\bullet)$ is the Dirac delta function. In view of the temporal and spatial homogeneities of fundamental solutions with respect to the singular point $\{t_0, \mathbf{x}_0\}$, we shall focus on the fundamental solution due to a point force located at the origin at $t = 0$. This fundamental solution satisfies

$$\nabla \bullet \mathbf{u} = 0, \quad \frac{\partial \mathbf{u}}{\partial t} + \mathbf{U} \bullet \nabla \mathbf{u} - \nabla \times [(\mathbf{\Omega} \times \mathbf{x}) \times \mathbf{u}] = -\nabla p + \nabla^2 \mathbf{u} + \mathbf{F}\delta(t)\delta(\mathbf{x}). \tag{6}$$

By means of Fourier transform in $\mathbf{x}$, $[\tilde{\mathbf{u}}, \tilde{p}](t, \mathbf{x}) = \int [\mathbf{u}, p](t, \mathbf{x}) e^{-i\mathbf{k} \bullet \mathbf{x}} d\mathbf{x}$, equation (6) can be expressed as $\mathbf{k} \bullet \tilde{\mathbf{u}} = 0$, $\frac{\partial \tilde{\mathbf{u}}}{\partial t} + (k^2 + i\mathbf{U} \bullet \mathbf{k})\tilde{\mathbf{u}} + \mathbf{k} \times [(\mathbf{\Omega} \times \tilde{\nabla}) \times \tilde{\mathbf{u}}] + i\tilde{p}\mathbf{k} = \mathbf{F}\delta(t)$, where $i = \sqrt{-1}$, $k = \|\mathbf{k}\|$, $\mathbf{k}$ is the vectorial wave number and a tilde above a term denotes its Fourier transform. From these equations, we find that $\tilde{p}$ is given by $\tilde{p} = -\frac{i}{k^2}\delta(t)\mathbf{F} \bullet \mathbf{k}$. The inverse Fourier transformation gives $p = \frac{\delta(t)\mathbf{F} \bullet \mathbf{x}}{4\pi r^3}$, $r = \|\mathbf{x}\|$. It is interesting to note that $p$ is independent of $\mathbf{U}$ and $\mathbf{\Omega}$. The fundamental solution of (6) is given by the following expression:

$$\mathbf{u} = \mathbf{\Phi}_3^+(\mathbf{\Phi}_2^+(\mathbf{\Phi}_1^+(\mathbf{a}))), \tag{7}$$

where $\mathbf{a} = \frac{1}{4\pi}\mathbf{F} \bullet (\mathbf{I}\nabla^2 - \nabla\nabla)f(t, \lambda)$, $f(t, \lambda) = -\frac{1}{2\sqrt{t}}\left[\frac{\text{erf}(\lambda)}{\lambda} - \frac{2}{\sqrt{\pi}}\right]$. In these expressions, we define $\mathbf{y}(t) = \int_0^t \mathbf{U}(\tau)d\tau$, $\theta_i(t) = \int_0^t \mathbf{\Omega}(\tau) \bullet \mathbf{e}_i \, d\tau$, for $i = 1, 2, 3$,

$$\eta = \|\mathbf{\Phi}_3^-(\mathbf{\Phi}_2^-(\mathbf{\Phi}_1^-(\mathbf{x}))) - \mathbf{y}\|, \quad \lambda = \frac{\eta}{2\sqrt{t}}. \tag{8}$$

$\boldsymbol{\Phi}_i^+$ and $\boldsymbol{\Phi}_i^-$ are orthogonal and linear operators, defined by $\boldsymbol{\Phi}_i^\pm(\mathbf{b}) = \mathbf{e}_i(\mathbf{b} \bullet \mathbf{e}_i) \mp (\mathbf{b} \times \mathbf{e}_i)\sin\theta_i \pm \mathbf{e}_i \times (\mathbf{b} \times \mathbf{e}_i)\cos\theta_i$ for $i = 1, 2, 3$ and the error function is given by $\mathrm{erf}(\xi) = \frac{2}{\sqrt{\pi}} \int_0^\xi e^{-\tau^2} d\tau$. The results obtained by Price & Tan [5], who only considered a special case in which the direction of $\boldsymbol{\Omega}(t)$ is fixed and parallel to that of $\mathbf{e}_1$ all the time, can be recovered from the generalized fundamental solution (7) by simply putting $\boldsymbol{\Phi}_2^\pm = \boldsymbol{\Phi}_3^\pm = \mathbf{I}$, an identity operator.

According to the theorem of splitting, the general solution (7) may be decomposed into two distinct types of waves, a longitudinal wave with irrotational velocity $\mathbf{u}_L$ and a transversal wave with uniform pressure and rotational velocity $\mathbf{u}_R$, where

$$\mathbf{u}_L = \mathbf{e}(\mathbf{a}_L \bullet \mathbf{e}) - (\mathbf{a}_L \times \mathbf{e})\sin\theta + \mathbf{e} \times (\mathbf{a}_L \times \mathbf{e})\cos\theta, \tag{9}$$

$$\mathbf{u}_R = \mathbf{e}(\mathbf{a}_R \bullet \mathbf{e}) - (\mathbf{a}_R \times \mathbf{e})\sin\theta + \mathbf{e} \times (\mathbf{a}_R \times \mathbf{e})\cos\theta, \tag{10}$$

$$\mathbf{a}_L = -\frac{1}{4\pi} \mathbf{F} \bullet \nabla\nabla f(t,\lambda), \quad \mathbf{a}_R = \frac{1}{4\pi} \mathbf{F} \nabla^2 f(t,\lambda). \tag{11}$$

This decomposition is unique up to specified conditions at infinity.

## 4  GENERALIZED UNSTEADY OSEENLET AND STOKESLET

For a purely translating body, i.e. $\boldsymbol{\Omega}(t) = \mathbf{O}$, equation (6) may be simplified to $\nabla \bullet \mathbf{u} = 0$, $\frac{\partial \mathbf{u}}{\partial t} + \mathbf{U} \bullet \nabla \mathbf{u} = -\nabla p + \nabla^2 \mathbf{u} + \mathbf{F}\delta(t)\delta(\mathbf{x})$, which is a generalized Oseen equation. The fundamental solution for unsteady translational motions can be derived by setting $\theta(t) = 0$ in equation (7). Thus $\mathbf{u} = \frac{1}{4\pi} \mathbf{F} \bullet (\mathbf{I}\nabla^2 - \nabla\nabla)f(t,\lambda)$, $p = \frac{\delta(t)\mathbf{F} \bullet \mathbf{x}}{4\pi r^3}$. To be consistent with the definition of the steady fundamental solution corresponding to infinite time $t$, we may define the unsteady fundamental solution $\mathbf{u}_H$ as $\mathbf{u}_H(t,\mathbf{x}) = \int_0^t \mathbf{u}(\tau,\mathbf{x})d\tau$, which corresponds to a Heaviside step change of the singular body force. Therefore

$$\mathbf{u}_H = \frac{1}{4\pi} \mathbf{F} \bullet (\mathbf{I}\nabla^2 - \nabla\nabla)g(t,r), \tag{12}$$

where

$$g(t,r) = \int_0^t f(\tau,\lambda)d\tau. \tag{13}$$

By analogy to the term Stokeslet, the fundamental solution (12) may be called a generalized unsteady Oseenlet. The steady Oseenlet can be obtained simply by letting time $t$ go to infinity in equation (13). If we set $\lambda = \frac{r}{2\sqrt{t}}$ in equation (13), we obtain the unsteady

Stokeslet kernel function $g_s(t,r) = -\frac{t}{r}\text{erf}\left(\frac{r}{2\sqrt{t}}\right) + \sqrt{\frac{t}{\pi}}\left(2 - e^{-\frac{r^2}{4t}}\right) + \frac{r}{2}\left[1 - \text{erf}\left(\frac{r}{2\sqrt{t}}\right)\right]$ and the steady Stokeslet kernel function $g_{s\infty}(r) = \frac{r}{2}$. The latter one is the same as the result of Chwang & Wu [3].

## 5 HYDRODYNAMIC FORCE ON A SPHERE TRANSLATING IMPULSIVELY IN A ROTATING VISCOUS FLOW

One of the most important objectives of the present paper is to obtain the hydrodynamic force acting on a solid body as a function of time $t$. Let us consider an unbounded unsteady Oseen flow with $\mathbf{U}(t) = \mathbf{e}_1 \text{H}(t)$ and $\mathbf{\Omega}(t) = \Omega \mathbf{e}_0 \text{H}(t)$ past a sphere of radius $R$ centered at the origin, where $\mathbf{e}_0$ and $\mathbf{e}_1$ are two constant unit vectors and $\text{H}(t)$ is Heaviside's step function. The diameter $2R$ is used as the characteristic length, $L = 2R$. For $t > 0$, the boundary conditions are then

$$\mathbf{u} = \mathbf{O} \text{ at } r = \frac{R_e}{2}, \quad \mathbf{u} \to \mathbf{e}_1 + \Omega \mathbf{e}_0 \times \mathbf{x} \text{ as } r \to \infty. \quad (14)$$

The flow due to the presence of the sphere may be obtained in terms of an unsteady Oseenlet and an unsteady potential doublet placed at the origin. Hence, the velocity is given by

$$\mathbf{u} = \mathbf{e}_1 + \Omega \mathbf{e}_0 \times \mathbf{x} + \int_0^t [\mathbf{e}_0(\mathbf{a}_f \bullet \mathbf{e}_0)] d\tau - \int_0^t \{[\sin(\Omega\tau) - \cos(\Omega\tau)\mathbf{e}_0 \times](\mathbf{a}_f \times \mathbf{e}_0)\} d\tau, \quad \text{where}$$

$\mathbf{a}_f = \frac{1}{4\pi}\mathbf{F} \bullet (\mathbf{I}\nabla^2 - \nabla\nabla) f(\tau,\lambda) + \mathbf{B} \bullet \nabla\nabla\frac{1}{r}$, $\mathbf{B}$ is the vectorial strength of the unsteady potential doublet. Obviously, $\mathbf{u}$ satisfies equation (6) and the boundary condition at infinity. Since $r = \frac{R_e}{2}$ on the sphere and $R_e$ is assumed small, we expand $\mathbf{a}_f$ for the small values of $r$ and obtain

$$\mathbf{u} \bullet \mathbf{e}_1 = 1 + \frac{2}{7}R_e\sqrt{\Omega(\mathbf{e}_0 \bullet \mathbf{e}_1)} - \frac{\mathbf{F} \bullet \mathbf{e}_1}{4\pi}\left[\frac{1}{2r} - \frac{G(t)}{4} + \frac{x}{4r} + \left(\frac{x}{2} - \frac{r^2 - x^2}{8}\right)\frac{x}{r^3}\right] - \frac{\mathbf{B} \bullet \mathbf{e}_1}{r^3}\left(1 - \frac{3x^2}{r^2}\right)$$
$$+ O(r \ln r + \Omega)$$

where $x = \mathbf{x} \bullet \mathbf{e}_1$ and $G(t) = \frac{2e^{-\frac{t}{4}}}{\sqrt{\pi t}} + \text{erf}\left(\frac{\sqrt{t}}{2}\right)$. Boundary condition (14) requires that

$$\mathbf{F} \bullet \mathbf{e}_1 = 3\pi R_e \left[1 + \frac{3}{16}R_e G(t)\right] \times \left[1 + \frac{2}{7}R_e\sqrt{\Omega(\mathbf{e}_0 \bullet \mathbf{e}_1)}\right] + O(R_e^3 \ln R_e + \Omega),$$

$$\mathbf{B} \bullet \mathbf{e}_1 = \frac{1}{32}R_e^3\left[1 + \frac{3}{16}R_e G(t)\right] \times \left[1 + \frac{2}{7}R_e\sqrt{\Omega(\mathbf{e}_0 \bullet \mathbf{e}_1)}\right] + O(R_e^5 \ln R_e + \Omega).$$ The drag comes only from the unsteady Oseenlet term, not from the term corresponding to the potential

doublet. As the dimensionless drag coefficient $C_D$ is normalized with respect to $\frac{1}{2}\rho U_0^2 \pi R^2$ instead of $\rho U_0^2 \left(\frac{L}{R_e}\right)^2$,

$$C_D = \frac{8(\mathbf{F} \bullet \mathbf{e}_I)}{\pi R_e^2} = \frac{24}{R_e}\left[1 + \frac{3}{16} R_e G(t)\right] \times \left[1 + \frac{2}{7} R_e \sqrt{\Omega(\mathbf{e}_0 \bullet \mathbf{e}_I)}\right] + O(R_e \ln R_e + \Omega).$$

This drag formula is a new result. The components of the angular velocity $\mathbf{\Omega}(t)$ in any direction perpendicular to $\mathbf{e}_I$ have no contribution to drag. In limiting cases, it agrees with the known results by Sano [6] and by Lovalenti & Brady [8] for pure translation ($\Omega = 0$) and by Childress [12] for steady motion ($t \to \infty$), as well as by the steady Oseen's theory [13], $C_{D_s} = \frac{24}{R_e}\left(1 + \frac{3}{16} R_e\right)$, which was experimentally verified to be accurate up to $R_e = 5$ approximately [14-16].

## 6 TWO-DIMENSIONAL GENERALIZED FUNDAMENTAL SOLUTIONS

The corresponding generalized fundamental solution for the two-dimensional problem is analogous to that for the three-dimensional case. Only the difference is discussed here for comparison. The generalized fundamental solution for the two-dimensional case is given by (7) with $\mathbf{a} = \frac{1}{4\pi} \mathbf{F} \bullet (\mathbf{I}\nabla^2 - \nabla\nabla) f_2(\lambda)$. In this expression, the kernel function $f_2(\lambda)$ does not depend on time $t$ explicitly and is a similarity function $f_2(\lambda) = 2\ln\lambda + E_I(\lambda^2) - \lambda^2 + \gamma$, where the similarity variable $\lambda$ is given by (8), $\gamma$ is Euler's constant and $E_I$ is the exponential integral, $E_I(\xi) = \int_\xi^\infty \frac{e^{-\tau}}{\tau} d\tau$. The solution may also be split into a longitudinal wave and a transversal wave given by equations (9) to (11) with $f(t,\lambda)$ replaced by $f_2(\lambda)$. The pressure for unsteady translational and rotational motion in an unbounded two-dimensional domain is given by $p_2 = \frac{\delta(t)\mathbf{F} \bullet \mathbf{x}}{2\pi r^2}$. The two-dimensional generalized unsteady Oseenlet may be defined as $\mathbf{u}_{H2}(t,\mathbf{x}) = \int_0^t \mathbf{u}(\tau,\mathbf{x}) d\tau$, which yields $\mathbf{u}_{H2} = \frac{1}{4\pi} \mathbf{F} \bullet (\mathbf{I}\nabla^2 - \nabla\nabla) g_2(t,r)$, $g_2(t,r) = \int_0^t f_2(\lambda) d\tau$. Consequently, the two-dimensional unsteady Stokeslet kernel function is $g_{2s}(t,r) = t\left[\ln\left(\frac{r^2}{4t}\right) + 1\right] + t\left(\gamma - e^{-\frac{r^2}{4t}}\right) + \left(t + \frac{r^2}{4}\right) E_I\left(\frac{r^2}{4t}\right) - \frac{r^2}{4}\ln t$ and the two-dimensional steady Stokeslet kernel function is $g_{2s\infty}(r) = -\frac{r^2}{4}\left[2\ln\left(\frac{r}{2}\right) + \gamma - 2\right]$.

# 7 HYDRODYNAMIC FORCE ON A CIRCULAR CYLINDER TRANSLATING IMPULSIVELY IN A ROTATING VISCOUS FLOW

We now consider an unsteady Oseen flow with $\mathbf{U}(t) = \mathbf{e}_1 \mathrm{H}(t)$ and $\mathbf{\Omega}(t) = \Omega \mathbf{e}_3$ past a circular cylinder of radius $R$ centered at the origin, where the constant unit vector $\mathbf{e}_3$ is perpendicular to the plane of the flow. The characteristic length is again defined as the diameter $2R$. The boundary conditions are $\mathbf{u} = \mathbf{O}$ at $r = \frac{R_e}{2}$, $\mathbf{u} \to \mathbf{e}_1 + \Omega \mathbf{e}_3 \times \mathbf{x}$ as $r \to \infty$. The solution consists of the uniform flow, a two-dimensional unsteady Oseenlet and a two-dimensional unsteady potential doublet located at the origin,

$$\mathbf{u} = \mathbf{e}_1 + \Omega \mathbf{e}_3 \times \mathbf{x} - \int_0^t \left[ (\mathbf{a}_{f2} \times \mathbf{e}_3) \sin(\Omega \tau) + \mathbf{a}_{f2} \cos(\Omega \tau) \right] d\tau,$$

where

$\mathbf{a}_{f2} = \frac{1}{4\pi} \mathbf{F} \bullet (\mathbf{I} \nabla^2 - \nabla \nabla) f_2(\lambda) + \mathbf{B} \bullet \nabla \nabla \ln r$. It is obvious that $\mathbf{u}$ satisfies equation (6) and the boundary condition at infinity. In view of $r = \frac{R_e}{2}$ on the circular cylinder and $R_e$ being small, we expand $\mathbf{u}$ for the small values of $r$ as

$$\mathbf{u} \bullet \mathbf{e}_1 \approx 1 - \frac{\mathbf{F} \bullet \mathbf{e}_1}{4\pi} S_0\left(\frac{t}{4}; \frac{r}{2}\right) - \frac{\mathbf{B} \bullet \mathbf{e}_1}{r^2}, \tag{15}$$

where the Shu function $S_n(t; \xi)$ is defined, similar to the Basset function or the modified Bessel function of the third kind, as

$$S_n(t; \xi) = \frac{1}{2} \left(\frac{\xi}{2}\right)^n \int_0^t \frac{e^{-\tau - \frac{\xi^2}{4\tau}}}{\tau^{n+1}} d\tau. \tag{16}$$

If we choose $\mathbf{F} \bullet \mathbf{e}_1 \approx \frac{8\pi}{1 + 2 S_0\left(\frac{t}{4}; \frac{R_e}{4}\right)}$, $\mathbf{B} \bullet \mathbf{e}_1 \approx \frac{R_e^2}{4\left[1 + 2 S_0\left(\frac{t}{4}; \frac{R_e}{4}\right)\right]}$, the right-hand side of equation (15) vanishes at $r = \frac{R_e}{2}$. As mentioned in Section 5, the drag comes only from the unsteady Oseenlet term. As the dimensionless drag coefficient is normalized with respect to $\rho U_0^2 R$ instead of $\rho U_0^2 \left(\frac{L}{R_e}\right)^2$, therefore $C_D = \frac{2(\mathbf{F} \bullet \mathbf{e}_1)}{R_e} \approx \frac{16\pi}{R_e \left[1 + 2 S_0\left(\frac{t}{4}; \frac{R_e}{4}\right)\right]}$. This drag formula is independent of $\Omega$. It agrees with the result of Nakanishi *et al.* [7], but the latter's expression is prohibitively complicated. In steady case, it also agrees with the steady Lamb's theory [17], $C_{D_s} = \dfrac{16\pi}{R_e \left[ 1 - 2\gamma - 2\ln\left(\dfrac{R_e}{8}\right) \right]}$, which was experimentally verified to be accurate up to $R_e = 1$ approximately [18].

## 12 CONCLUSIONS

A number of new closed-form generalized fundamental solutions have been derived in the present paper for general time-dependent linearized viscous flows. The combination of these generalized fundamental solutions can provide solutions to a wide variety of unsteady flow problems. They also provide a comprehensive framework for the singularity method in dealing with unsteady linearized motions, especially for the flow associated with unsteady translational and rotational motions. It is demonstrated that the new generalized fundamental solutions can be used to calculate the time-evolution of drag coefficients for a sphere and a circular cylinder. New results are obtained for a rotating viscous flow past an impulsively moving sphere and an impulsively moving circular cylinder. A suitable arrangement of these new generalized fundamental solutions may produce solutions for a general unsteady flow past a body of arbitrary shape.